\magnification \magstep1
\raggedbottom
\openup 1\jot
\voffset6truemm
\leftline {\bf NON-FUCHSIAN SINGULARITIES IN QUANTUM MECHANICS}
\vskip 1cm
\leftline {$\; \; \; \; \; \; \; \; \; \; \; \; \; \; \; \; \;$
{\bf Giampiero Esposito}}
\vskip 0.3cm
\leftline {$\; \; \; \; \; \; \; \; \; \; \; \; \; \; \; \; \;$
{\it INFN, Sezione di Napoli}}
\leftline {$\; \; \; \; \; \; \; \; \; \; \; \; \; \; \; \; \;$
{\it Mostra d'Oltremare Padiglione 20}}
\leftline {$\; \; \; \; \; \; \; \; \; \; \; \; \; \; \; \; \;$
{\it 80125 Napoli, Italy}}
\leftline {$\; \; \; \; \; \; \; \; \; \; \; \; \; \; \; \; \;$ and}
\leftline {$\; \; \; \; \; \; \; \; \; \; \; \; \; \; \; \; \;$
{\it Universit\`a di Napoli Federico II}}
\leftline {$\; \; \; \; \; \; \; \; \; \; \; \; \; \; \; \; \;$
{\it Dipartimento di Scienze Fisiche}}
\leftline {$\; \; \; \; \; \; \; \; \; \; \; \; \; \; \; \; \;$
{\it Complesso Universitario di Monte S. Angelo}}
\leftline {$\; \; \; \; \; \; \; \; \; \; \; \; \; \; \; \; \;$
{\it Via Cintia, Edificio G}}
\leftline {$\; \; \; \; \; \; \; \; \; \; \; \; \; \; \; \; \;$
{\it 80126 Napoli, Italy}}
\vskip 1cm
\noindent
A modification of the
spiked harmonic oscillator is studied 
in the case for which the perturbation
potential contains both an inverse power and a linear term. It
is then possible to evaluate trial functions by solving an integral
equation due to the occurrence of the linear term. The general form
of such integral equation is obtained by using a Green-function
method, and adding a modified Bessel function 
of second kind which solves an homogeneous
problem with Dirichlet boundary condition at the origin.
\vskip 100cm
\leftline {\bf 1. INTRODUCTION}
\vskip 0.3cm
\noindent
In the literature on quantum-mechanical problems, an important 
role is played by the ``spiked'' harmonic oscillator in three 
spatial dimensions. This is a system where 
the radial part $\psi(r)$ of the
wave function is ruled by the Hamiltonian operator [1--4]
$$
{\widetilde H}(\alpha,\lambda)\equiv -{d^{2}\over dr^{2}}
+r^{2}+{l(l+1)\over r^{2}}+{\lambda \over r^{\alpha}},
\eqno (1.1)
$$
in the sense that ${\widetilde H}(\alpha,\lambda)$ acts on 
$\varphi(r) \equiv r \psi(r)$.
In the $s$-wave case (i.e. when the angular momentum quantum
number $l$ vanishes) ${\widetilde H}(\alpha,\lambda)$ 
reduces to 
$$
H(\alpha,\lambda) \equiv -{d^{2}\over dr^{2}}+r^{2}
+{\lambda \over r^{\alpha}} \equiv H(\alpha,0)+\lambda V,
\eqno (1.2)
$$
where $H(\alpha,0)$ is formally equal to the simple harmonic oscillator
Hamiltonian, $r$ belongs to the interval $[0,\infty]$ and
$V \equiv r^{-\alpha}$. For any fixed value of $\lambda$, the
potential term in (1.2) diverges as $r \rightarrow 0$ in such a
way that the operator $H(\alpha,\lambda)$ acts on wave functions
which vanish at the origin:
$$
\varphi(0)=0.
\eqno (1.3)
$$
More precisely, the imposition of Eq. (1.3) is necessary since
not all functions in the domain of $H(\alpha,0)$ are in the
domain of $V$ [2]. Thus, when
$\lambda \rightarrow 0$ and $\alpha$ is fixed, the
operator $H(\alpha,\lambda)$ converges to an operator formally
equal to the unperturbed operator 
$H_{0} \equiv -{d^{2}\over dr^{2}}+r^{2}$,
but supplemented by the Dirichlet boundary condition (1.3) for
all functions in its domain. This means that the unperturbed 
operator $H_{0}$, for which the 
boundary condition (1.3) is not necessary,
differs from the limiting operator $H(\alpha,0)$, for which Eq.
(1.3) is instead necessary to characterize the domain.

The full potential in (1.1) or (1.2) inherits the name ``spiked''
from a pronounced peak near the origin for $\lambda >0$, and its
consideration is suggested by many concrete problems in chemical,
nuclear and particle physics. Some important results within this
framework are as follows.
\vskip 0.3cm
\noindent
(i) Development of singular perturbation theory, with application
to the small-$\lambda$ expansion of the ground-state energy [1].
\vskip 0.3cm
\noindent
(ii) A variational method has been successfully applied to a
large-coupling perturbative calculation of the ground-state
energy [2].
\vskip 0.3cm
\noindent
(iii) Weak-coupling perturbative analysis [3], and its relation with
the strong coupling regime for $\alpha$ in the neighbourhood
of $\alpha=2$.
\vskip 0.3cm
\noindent
(iv) A non-perturbative but absolutely convergent algorithm for the
evaluation of eigenfunctions [4].
\vskip 0.3cm
Note now that for $\alpha >2$ the Hamiltonian operators (1.1) or 
(1.2) lead to a non-Fuchsian singularity [5] at $r=0$ of the 
stationary Schr\"{o}dinger equation, because their potential term
has a pole of order $>2$ therein. Section $2$ outlines the
method developed by Harrell [1] for dealing with such singularities
in a perturbative analysis. A non-trivial extension is studied
in Sec. $3$, i.e. a model where the perturbation potential
contains both an inverse power and a linear term. Concluding
remarks are presented in Sec. 4.
\vskip 0.3cm
\leftline {\bf 2. THE HARRELL METHOD}
\vskip 0.3cm
\noindent
The method developed by Harrell relies on the choice of suitable
trial functions for self-adjoint operators (i.e. normalized
vectors in their domain) and on the following lemma [1]:
\vskip 0.3cm
\noindent
{\bf Lemma.}
If $\psi_{\lambda}$ is a trial function for the self-adjoint 
operator $T+\lambda T'$, where both $T$ and $T'$ are self-adjoint
and $E(0)$ is an isolated, non-degenerate stable eigenvalue of
$T$, and $E(\lambda)$ is a continuous function such that the
scalar product $\Bigr(\psi_{\lambda},[T+\lambda T' -E(\lambda)]
\psi_{\lambda}\Bigr)$ tends to $0$ as $\lambda$ tends to $0$, and
$$
\left \| [T+\lambda T' -E(\lambda)] \psi_{\lambda} 
\right \|={\rm o} \left(\sqrt{(\psi_{\lambda},
[T+\lambda T' -E(\lambda)]\psi_{\lambda})}\right),
\eqno (2.1)
$$
then the eigenvalue of $T+\lambda T'$ which converges 
to $E(0)$ is
$$
E(\lambda)=
\Bigr(\psi_{\lambda},[T+\lambda T']\psi_{\lambda}\Bigr)
+{\rm O} \left( \left \| [T+\lambda T'-E(\lambda)]
\psi_{\lambda} \right \|^{2} \right).
\eqno (2.2)
$$

In a one-dimensional example, hereafter chosen for simplicity,
let the full Hamiltonian be $H_{0}+\lambda V$, where
$$
H_{0}\equiv -{d^{2}\over dx^{2}}+x^{2},
\eqno (2.3)
$$
and
$$
V(x)\equiv x^{-\alpha}.
\eqno (2.4)
$$
If $\alpha=4$, on denoting by $u_{i}$ the unperturbed eigenstates,
a trial function can be chosen in the form [1]
$$
\psi(x)=W(x;\lambda)u_{i}(x),
\eqno (2.5)
$$
where (inspired by the JWKB approximation)
$$
W(x;\lambda)=N(\lambda){\rm exp} \left(-\int_{x}^{\infty}
\sqrt{\lambda V(\xi)}d\xi \right)
=N(\lambda){\rm exp} \left(-{\sqrt{\lambda}\over x}\right),
\eqno (2.6)
$$
with $N(\lambda)$ a normalization factor that approaches $1$
as $\lambda$ tends to $0$. One then finds the formula [1]
$$
[H_{0}-E_{i}+\lambda V]W u_{i}=2W{\sqrt{\lambda}\over x^{2}}
\left({1\over x}-{d\over dx}\right)u_{i}.
\eqno (2.7)
$$
By virtue of Eq. (2.7) and of the Lemma at the beginning of
this section one obtains the following formula for the energy
eigenvalues of the spiked harmonic oscillator in one
dimension with $\alpha=4$:
$$
E_{i}(\lambda)=E_{i}(0)+2 \left(u_{i},x^{-2}
\left({1\over x}-{d\over dx}\right) u_{i} \right)
\sqrt{\lambda}+{\rm O}(\lambda).
\eqno (2.8)
$$
When $\alpha \not = 4$, one can use the identity (hereafter
$W_{\alpha}$ replaces $W$)
$$
{d^{2}\over dx^{2}}(W_{\alpha}u_{i})=W_{\alpha}u_{i}''
+2W_{\alpha}'u_{i}'+W_{\alpha}''u_{i},
\eqno (2.9)
$$
which implies [1]
$$
[H_{0}-E_{i}+\lambda V]W_{\alpha}u_{i}=
\left[-{d^{2}W_{\alpha}\over dx^{2}}-2{dW_{\alpha}\over dx}
{d\over dx}+\lambda V W_{\alpha}\right]u_{i}.
\eqno (2.10)
$$
The idea is now to choose $W_{\alpha}$ in such a way that the
action of $[H_{0}-E_{i}+\lambda V]$ on $W_{\alpha}u_{i}$ involves
again the action of the operator $\left({1\over x}
-{d\over dx}\right)$ on $u_{i}$ (see Eq. (2.7)). For this purpose,
Harrell imposed the differential equation
$$
\left[{d^{2}\over dx^{2}}+{2\over x}{d\over dx}
-{\lambda \over x^{\alpha}}\right]W_{\alpha}=0,
\eqno (2.11)
$$
so that Eq. (2.10) reduces indeed to (cf. Eq. (2.7))
$$
[H_{0}-E_{i}+\lambda V]W_{\alpha}u_{i}
=2{dW_{\alpha}\over dx}\left({1\over x}-{d\over dx}\right)u_{i}.
\eqno (2.12)
$$
On defining
$$
\nu \equiv {1\over (\alpha-2)},
\eqno (2.13)
$$
the solution of Eq. (2.11) can be expressed in the form
$$
W_{\alpha}(x;\lambda)={2\nu^{\nu}\lambda^{\nu \over 2}\over
\Gamma(\nu)}x^{-{1\over 2}}
K_{\nu}(2\nu \sqrt{\lambda}x^{-{1\over 2\nu}}).
\eqno (2.14)
$$
The energy eigenvalues of the spiked oscillator in one dimension
with $\alpha \geq 4$ are then found to be [1]
$$
E_{i}(\lambda)=E_{i}(0)+2{\Gamma(1-\nu)\over \Gamma(1+\nu)}\nu^{2\nu}
\left(u_{i},x^{-2}\left({1\over x}-{d\over dx}\right)u_{i}\right)
\lambda^{\nu}+{\rm O}(\lambda^{2\nu}).
\eqno (2.15)
$$
\vskip 0.3cm
\leftline {\bf 3. EXTENSION TO OTHER SINGULAR POTENTIALS}
\vskip 0.3cm
\noindent
When the perturbation potential is not an inverse power, Eq. (2.11)
is replaced by an equation which cannot generally be solved. Harrell
has however shown that, if $V$ is bounded away from $x=0$ and lies
in between $x^{-\alpha}$ and $x^{-\beta}$, with 
$0 < \alpha < \beta$, then the effect of $V$ on the eigenvalues
is not essentially different [1].

It therefore appears of interest to study cases in which $V$ is
not (a pure) inverse power, but does not obey the restrictions 
considered in Sec. 5 of Ref. [1]. For this purpose, we assume
that (cf. (2.4))
$$
V(x)\equiv x^{-\alpha}+\kappa x.
\eqno (3.1)
$$
By doing so we study a model that reduces to the spiked oscillator
in the neighbourhood of the origin, whereas at large $x$ it 
approaches an oscillator perturbed by a ``Stark-like'' term.
The two terms in (3.1) are separately well studied in the 
literature, so that their joint effect provides a well motivated
departure from the scheme studied in Sec. 5 of Ref. [1].

The exact solution of the counterpart of Eq. (2.11) when (3.1)
holds is not available in the literature to our knowledge. For this
purpose we study an integral equation, whose construction
is as follows. On assuming the validity of Eq. (3.1) for the
perturbation potential, Eq. (2.11) is replaced by the 
inhomogeneous equation
$$
LW_{\alpha}(x;\lambda)=f_{\alpha}(x;\lambda),
\eqno (3.2)
$$
where
$$
L \equiv {d^{2}\over dx^{2}}+{2\over x}{d\over dx}
-{\lambda \over x^{\alpha}},
\eqno (3.3)
$$
$$
f_{\alpha}(x;\lambda) \equiv 
\lambda \kappa x W_{\alpha}(x;\lambda).
\eqno (3.4)
$$
Since $V(x)$ approaches $x^{-\alpha}$ as $x \rightarrow 0$, we
require again the boundary condition studied in Ref. [1], i.e.
$$
W_{\alpha}(0)=0.
\eqno (3.5)
$$
We are here following an approach similar to the one leading to
an integral equation for scattering problems, where the 
right-hand side of the stationary Schr\"{o}dinger equation
involves again the unknown function:
$$
(\bigtriangleup + k^{2})\psi_{k}({\vec x})
={2m\over {\hbar}^{2}}V({\vec x})\psi_{k}({\vec x}).
$$
In our problem, following Ref. [1], we study one-dimensional
Schr\"{o}dinger operators with Dirichlet boundary conditions
at $0$ only on the space $L^{2}([0,\infty))$, to remove the
degeneracy resulting from the decoupling of the two half-lines
$(-\infty,0)$ and $(0,\infty)$ (only odd eigenfunctions of the
ordinary harmonic oscillator obey the Dirichlet condition at
$0$). Moreover, to be able to use all known results on
one-dimensional boundary-value problems on closed intervals 
of the real line, we first work on the interval $(0,b)$ and
then take the limit as $b \rightarrow \infty$. More precisely,
we start from a problem whose Green function 
$G(x,\xi;\lambda)$ satisfies the equation
$$
LG=0 \; \; {\rm for} \; \; x \in (0,\infty) \; \; {\rm and}
\; \; \xi \in (0,\infty),
\eqno (3.6)
$$
the boundary condition
$$
G(0,\xi;\lambda)=0,
\eqno (3.7)
$$
the summability condition (in that the integral of $G$ over a
closed interval of values of $\lambda$ yields a square-integrable
function of $x$)
$$
G(x,\xi;\lambda) \in L_{2}^{(c)}(0,\infty),
\eqno (3.8)
$$
and the continuity conditions
$$
\lim_{x \to \xi^{+}}G(x,\xi;\lambda)=
\lim_{x \to \xi^{-}}G(x,\xi;\lambda),
\eqno (3.9)
$$
$$
\lim_{x \to \xi^{+}}{\partial G \over \partial x}
-\lim_{x \to \xi^{-}}{\partial G \over \partial x}=1.
\eqno (3.10)
$$
As is shown on page 442 of Ref. [5], $G(x,\xi;\lambda)$ is
recovered on studying first the Green function 
$G_{b}(x,\xi;\lambda)$ for a regular problem on the interval
$(0,b)$, and then taking the limit as $b \rightarrow \infty$, i.e.
$$
\lim_{b \to \infty}G_{b}(x,\xi;\lambda)=G(x,\xi;\lambda).
\eqno (3.11)
$$
Such a relation holds independently of the boundary condition
imposed on $G_{b}(x,\xi;\lambda)$ at $x=b$ [5]. With this
understanding, the full solution of the inhomogeneous equation
(3.2) with boundary condition (3.5) is given by ($\gamma$ being
a constant)
$$ \eqalignno{
\; & W_{\alpha}(x;\lambda)=\gamma 
{2\nu^{\nu}\lambda^{\nu \over 2}\over \Gamma(\nu)}x^{-{1\over 2}}
K_{\nu}(2\nu \sqrt{\lambda}x^{-{1\over 2\nu}}) \cr
&+\lim_{b \to \infty} \lambda \kappa \int_{0}^{b}
G_{b}(x,\xi;\lambda)\xi W_{\alpha}(\xi;\lambda)d\xi,
&(3.12)\cr}
$$
where the first term on the right-hand side of (3.12) is the regular
solution (2.14) of the homogeneous equation $LW_{\alpha}=0$. The
Green function $G_{b}(x,\xi;\lambda)$ has to obey the 
differential equation [5]
$$
L G_{b}=0 \; \; {\rm for} \; \; x \in (0,\xi)
\; \; {\rm and} \; \; x \in (\xi,b),
\eqno (3.13)
$$
the homogeneous boundary conditions [5]
$$
G_{b}(0,\xi;\lambda)=0,
\eqno (3.14)
$$
$$
G_{b}(b,\xi;\lambda)=0,
\eqno (3.15)
$$
and the continuity conditions [5]
$$
\lim_{x \to \xi^{+}}G_{b}(x,\xi;\lambda)
=\lim_{x \to \xi^{-}}G_{b}(x,\xi;\lambda),
\eqno (3.16)
$$
$$
\lim_{x \to \xi^{+}}{\partial G_{b}\over \partial x}
-\lim_{x \to \xi^{-}}{\partial G_{b}\over \partial x}=1.
\eqno (3.17)
$$
To obtain the explicit form of $G_{b}(x,\xi;\lambda)$ one has to
consider a non-trivial solution $u_{0}(x;\lambda)$ of the 
homogeneous equation $Lu=0$ satisfying $u(0)=0$, and a non-trivial
solution $u_{b}(x;\lambda)$ of $Lu=0$ satisfying $u(b)=0$. By 
virtue of (3.13)--(3.15) one then finds [5]
$$
G_{b}(x,\xi;\lambda)=A(\xi;\lambda)u_{0}(x;\lambda) 
\; \; {\rm if} \; \; x \in (0,\xi),
\eqno (3.18)
$$
$$
G_{b}(x,\xi;\lambda)=B(\xi;\lambda)u_{b}(x;\lambda) 
\; \; {\rm if} \; \; x \in (\xi,b),
\eqno (3.19)
$$
where $u_{0}$ and $u_{b}$ are independent. The
continuity conditions (3.16) and (3.17) imply that $A(\xi;\lambda)$ 
and $B(\xi;\lambda)$ are obtained by solving the inhomogeneous system
$$
A(\xi;\lambda)u_{0}(\xi;\lambda)-B(\xi;\lambda)u_{b}(\xi;\lambda)=0,
\eqno (3.20)
$$
$$
B(\xi;\lambda)u_{b}'(\xi;\lambda)-A(\xi;\lambda)u_{0}'(\xi;\lambda)=1,
\eqno (3.21)
$$
which yields
$$
A(\xi;\lambda)={u_{b}(\xi;\lambda)\over 
\Omega(u_{0},u_{b};\xi;\lambda)},
\eqno (3.22)
$$
$$
B(\xi;\lambda)={u_{0}(\xi;\lambda)\over 
\Omega(u_{0},u_{b};\xi;\lambda)},
\eqno (3.23)
$$
where $\Omega$ is the Wronskian of $u_{0}$ and $u_{b}$.
We now recall the Abel formula for $\Omega$, according to which [5]
$$
\Omega(u_{0},u_{b};x;\lambda)=C(\lambda) e^{-v(x)},
\eqno (3.24)
$$
where $C(\lambda)$ is a constant and, for the operator $L$ in (3.3),
$v$ is a particular solution of the equation [5]
$$
{dv\over dx}={2\over x},
\eqno (3.25)
$$
i.e.
$$
v(x)=\log x^{2},
\eqno (3.26)
$$
which implies that
$$
\Omega={C \over x^{2}}.
\eqno (3.27)
$$
Thus, on defining as usual $x_{<} \equiv {\rm min}(x,\xi),
x_{>} \equiv {\rm max}(x,\xi)$, Eqs. (3.18), (3.19), (3.22),
(3.23) and (3.27) imply that the Green function is expressed by
$$
G_{b}(x,\xi;\lambda)={\xi^{2}\over C(\lambda)}
u_{0}(x_{<};\lambda)u_{b}(x_{>};\lambda).
\eqno (3.28)
$$
The integral equation (3.12) for $W_{\alpha}$ becomes therefore
$$ \eqalignno{
\; & W_{\alpha}(x;\lambda)=\gamma 
{2\nu^{\nu}\lambda^{\nu \over 2}\over \Gamma(\nu)}x^{-{1\over 2}}
K_{\nu}(2\nu \sqrt{\lambda}x^{-{1\over 2\nu}}) \cr
&+\lim_{b \to \infty}{\lambda \kappa \over C}
\left[u_{b}(x;\lambda)\int_{0}^{x}u_{0}(\xi;\lambda)
\xi^{3}W_{\alpha}(\xi;\lambda)d\xi \right . \cr
& \left . + u_{0}(x;\lambda)\int_{x}^{b}u_{b}(\xi;\lambda)
\xi^{3} W_{\alpha}(\xi;\lambda)d\xi \right].
&(3.29)\cr}
$$

In Eq. (3.29), $u_{0}(x;\lambda)$ and $u_{b}(x;\lambda)$ can be
chosen to be of the form
$$
u_{0}(x;\lambda)=C_{0}(\nu)x^{-{1\over 2}}
K_{\nu}(2\nu \sqrt{\lambda}x^{-{1\over 2\nu}}),
\eqno (3.30)
$$
$$
u_{b}(x;\lambda)=C_{b}(\nu)x^{-{1\over 2}}
I_{\nu}(2\nu \sqrt{\lambda}x^{-{1\over 2\nu}}),
\eqno (3.31)
$$
in agreement with the homogeneous Dirichlet conditions at $0$ 
and at $b$, respectively. Before taking the limit as 
$b \rightarrow \infty$ in Eq. (3.29) we can now regard the
perturbation parameter $\lambda$ as an eigenvalue. We are
therefore studying, for finite $b$, a Fredholm integral
equation of second kind, whose general form is (the parameter
$a$ vanishes in our problem)
$$
\varphi(s)=f(s)+\lambda \int_{a}^{b}K(s,t)\varphi(t)dt.
\eqno (3.32)
$$
If $\lambda$ is an eigenvalue, a necessary and sufficient condition
for the existence of solutions of Eq. (3.32) is that, for any
solution $\chi$ of the equation
$$
\chi(s)=\lambda \int_{a}^{b}K(t,s)\chi(t)dt,
\eqno (3.33)
$$
the known term $f(s)$ should satisfy the condition [6]
$$
\int_{a}^{b}f(s)\chi(s)ds=0.
\eqno (3.34)
$$
In our problem, $f$ is the first term on the right-hand side of
Eq. (3.29), the kernel $K$ is given by $G_{b}$ in (3.28), and
Eq. (3.34) provides a powerful operational criterion. Indeed,
one might try to use directly the theory of integral equations
[6] on the interval $(0,\infty)$ instead of the limiting procedure
in Eq. (3.29), but the necessary standard of rigour goes beyond
our present capabilities.
\vskip 0.3cm
\leftline {\bf 4. CONCLUDING REMARKS}
\vskip 0.3cm
\noindent
In the present letter we have exploited 
the fact that if a spiked harmonic
oscillator is modified by the addition of a linear term, the full
perturbation potential may be seen as consisting of an inverse
power plus a term linear in the independent variable. It is then
possible to evaluate the function $W(x;\lambda)$ occurring in the
trial function (2.5) by solving the inhomogeneous equation (3.2),
which leads to the integral equation (3.29). This is involved, but
leads in principle to a complete calculational scheme. Note
that, if one tries to combine the term $x^{2}$ in the unperturbed
Hamiltonian with the linear term in the perturbation potential
(3.1), one eventually moves the singular point away from the
origin, whereas the spiked oscillator is (normally) studied by
looking at the singular point at the origin.

It would be rather interesting, as a subject for further research,
to consider suitable changes of independent variable in the
investigation of non-Fuchsian singularities. For example, given
the Hamiltonian operator 
$$
H \equiv -{d^{2}\over dr^{2}}+{b\over r^{2}}
+{a\over r^{p}},
\eqno (4.1)
$$
if one defines the new independent variable (cf. page 971
of Ref. [7])
$$
\rho \equiv r^{\gamma},
\eqno (4.2)
$$
for a suitable parameter $\gamma$, the stationary Schr\"{o}dinger
equation becomes
$$
\left[{d^{2}\over d\rho^{2}}+\left(1-{1\over \gamma}\right)
{1\over \rho}{d\over d\rho}+{1\over \gamma^{2}\rho^{2}}
\left(E\rho^{2\over \gamma}-a \rho^{{(2-p)\over \gamma}}-b
\right) \right]\varphi(\rho)=0.
\eqno (4.3)
$$
By construction, the larger is $\gamma$, the more Eq. (4.3) tends
to its Fuchsian limit 
$$
\left[{d^{2}\over d\rho^{2}}+\left(1-{1\over \gamma}\right)
{1\over \rho}{d\over d\rho}
+{(E-a-b)\over \gamma^{2}}{1\over \rho^{2}}\right]
\varphi(\rho)=0,
\eqno (4.4)
$$
for all values of $\rho$. This remark can be made precise 
by defining
$$
\varepsilon \equiv {1\over \gamma},
\eqno (4.5)
$$
$$
F(\varepsilon) \equiv \varepsilon^{2}\Bigr(E \rho^{2\varepsilon}
-a \rho^{(2-p)\varepsilon}-b \Bigr),
\eqno (4.6)
$$
and considering the asymptotic expansion at small $\varepsilon$
(and hence large $\gamma$)
$$
F(\varepsilon) \sim \varepsilon^{2} \Bigr[E-a-b
+\varepsilon (2E-(2-p)a)\log \rho +{\rm O}(\varepsilon^{2})
\Bigr].
\eqno (4.7)
$$
The first ``non-Fuchsian correction'' of the limiting equation
(4.4) is therefore
$$ 
\left \{ {d^{2}\over d\rho^{2}}
+(1-\varepsilon) {1\over \rho} {d\over d\rho} 
+ {1\over \rho^{2}}\Bigr[\varepsilon^{2}
((E-a-b)+\varepsilon(2E-(2-p)a)\log \rho)\Bigr]
\right \} \varphi(\rho)=0.
\eqno (4.8) 
$$
Interestingly, logarithmic terms in the potential can be therefore
seen to result from a sequence of approximations relating Eq. (4.3)
to its Fuchsian limit (4.4). Moreover, all equations with Fuchsian
singularities like Eq. (4.4) might be seen as non-trivial limits
of stationary Schr\"{o}dinger equations with non-Fuchsian 
singular points. It remains to be seen whether such properties
can be useful in the investigation of the topics discussed in
the previous sections.

Another topic for further research is the Schr\"{o}dinger equation
for perturbed stationary states of an isotropic oscillator in three
dimensions written in the form
$$
\left[{d^{2}\over dr^{2}}+k^{2}-\mu^{2}r^{2}
-{l(l+1)\over r^{2}}-S(r)\right]\varphi(r)=0,
\eqno (4.9)
$$
where, having set (here $\mu \equiv {m\omega \over \hbar}$)
$$
V(r)\equiv {2m\over \hbar^{2}}U(r)=\mu^{2}r^{2}+S(r),
\eqno (4.10)
$$
the function $S$ represents the ``singular'' part of the potential
according to our terminology. We look for exact solutions of 
Eq. (4.9) which can be written as 
$$
\varphi(r)=A(r)e^{B(r)}e^{-{\mu r^{2}\over 2}}.
\eqno (4.11)
$$
The second exponential in (4.11) takes into account that, at
large $r$, the term $\lambda^{2}r^{2}$ dominates over all other
terms in the potential (including, of course, 
${l(l+1)\over r^{2}}$), and has not been absorbed into $B(r)$
for later convenience. It is worth stressing that Eq. (4.11)
is not a JWKB ansatz but rather a convenient factorization of
the exact solution of Eq. (4.9). We determine $B(r)$ from a
non-linear equation by straightforward integration (see below),
while the corresponding second-order equation for $A$ is
rather involved.

Indeed, insertion of (4.11) into Eq. (4.9) leads to
$$
\left \{ {d^{2}\over dr^{2}}+2(B'-\mu r){d\over dr}
+\left[k^{2}-\mu-{l(l+1)\over r^{2}}-2\mu r B'
+B''+{B'}^{2}-S(r)\right]\right \}A(r)=0.
\eqno (4.12)
$$
To avoid having coefficients of this equation which depend in a
non-linear way on $B$ we {\it choose} the function $B$ so that
$$
{B'}^{2}-S(r)=0,
\eqno (4.13)
$$
which implies (up to a sign, here implicitly absorbed into
the square root)
$$
B(r)=\int \sqrt{S(r)} \; dr.
\eqno (4.14)
$$
Hence one finds the following second-order equation for the
function $A$:
$$
\left \{ {d^{2}\over dr^{2}}+2(\sqrt{S}-\mu r){d\over dr}
+\left[k^{2}-\mu-{l(l+1)\over r^{2}}-2\mu r \sqrt{S}
+{S'\over 2\sqrt{S}}\right]\right \}A(r)=0.
\eqno (4.15)
$$
It should be stressed that the step leading to Eq. (4.13) is
legitimate but not compelling. For each choice of $B(r)$ there
will be a different equation for $A(r)$, but in such a way that
$\varphi(r)$ remains the same (see (4.11)). Unfortunately,
Eq. (4.15) remains too difficult, as far as we can see.
\vskip 0.3cm
\noindent
{\bf Acknowledgments.} This work has been partially supported
by PRIN97 ``Sintesi.''
\vskip 0.3cm
\leftline {\bf REFERENCES}
\vskip 0.3cm
\item {[1]}
E. M. Harrell, {\it Ann. Phys. (N.Y.)} {\bf 105}, 379 (1977).
\item {[2]}
V. C. Aguilera--Navarro, G. A. Est\'evez and R. Guardiola,
{\it J. Math. Phys.} {\bf 31}, 99 (1990).
\item {[3]}
V. C. Aguilera--Navarro and R. Guardiola, {\it J. Math. Phys.}
{\bf 32}, 2135 (1991).
\item {[4]}
H. G. Miller, {\it J. Math. Phys.} {\bf 35}, 2229 (1994).
\item {[5]}
I. Stakgold, {\it Green's Functions and Boundary Value Problems}
(John Wiley, New York, 1979).
\item {[6]}
V. I. Smirnov, {\it Course in Higher Mathematics, Vol. IV, Part I}
(Mir, Moscow, 1980).
\item {[7]}
I. S. Gradshteyn and I. M. Ryzhik, {\it Table of Integrals, 
Series and Products} (Academic, New York, 1965).

\bye